\documentclass[reprint,
superscriptaddress,
 amsmath,amssymb,
 aps,
pra,
]{revtex4-1}
\usepackage{dsfont}

\usepackage{nicefrac}
\usepackage{braket}
\usepackage{xcolor}
\usepackage{graphicx}
\usepackage{tikz}
\usepackage{dcolumn}
\usepackage{bm}
\usepackage{hyperref}
\usepackage[caption=false]{subfig}
\usepackage[normalem]{ulem}
\usepackage{float}
\usepackage{soulutf8}
\usepackage{algorithm}
\usepackage{algpseudocode}
\usepackage{algorithmicx}
\usepackage{cancel}

\usepackage[hang,flushmargin]{footmisc}
\setstcolor{red}

\begin{document}

\preprint{APS/123-QED}

\title{Mitigating Phase Correlations in Quantum Key Distribution Using Path-Selection Modulation}

\author{Amita Gnanapandithan}
\affiliation{Department of Electrical and Computer Engineering, University of Toronto, Toronto, Canada}

\author{Li Qian}
\affiliation{Department of Electrical and Computer Engineering, University of Toronto, Toronto, Canada}
\affiliation{Center for Quantum Information and Quantum Control, University of Toronto, Toronto, Canada}

\author{Hoi-Kwong Lo}
\affiliation{Department of Electrical and Computer Engineering, University of Toronto, Toronto, Canada}
\affiliation{Center for Quantum Information and Quantum Control, University of Toronto, Toronto, Canada}
\affiliation{Department of Physics, University of Toronto, Toronto, Canada}
\affiliation{Department of Physics, National University of Singapore, Singapore}

\date{\today}

\begin{abstract}
    Phase correlations are an under-explored vulnerability in QKD. Here, we present an experimental and simulated characterization of correlations arising from electro-optic phase encoding, over repetition rates up to the GHz level. To mitigate this vulnerability (and all side channels arising from active phase modulators), we propose a ``path-selection modulation'' source that eliminates the need for active phase modulation altogether. Encoding is achieved by randomly selecting between multiple paths, each path corresponding to one of the desired encoded states. Phase randomization is achieved using gain-switching.  We characterize this source at a clock rate of 1 GHz. 
\end{abstract}
 
\maketitle

\section{Introduction}
\label{sec:intro}
Quantum key distribution (QKD) is a technique enabling two parties to securely share a secret key by leveraging the postulates of quantum mechanics \cite{XuRMP2020}. This process involves encoding information into a sequence of quantum optical states, often utilizing polarization or phase states. Electro-optic phase modulators (PM) are commonly used in this encoding process.

Despite the theoretical security guarantees of QKD, practical implementations often fall short due to various imperfections. These imperfections manifest as side channels that can potentially leak information to an eavesdropper \cite{Kurtsiefer2001,Makarov2006,Brassard2000,Sun2015,Fung2007}. While the measurement-device-independent QKD (MDI QKD) protocol eliminates detector side channels \cite{Lo2012}, source side channels continue to pose a significant risk \cite{Brassard2000,Sun2015,Fung2007}.

The recent drive towards high-speed QKD sources operating at GHz clock rates has unveiled new source side channels that demand attention. Bandwidth-limited modulation has been shown to introduce hidden multidimensional modulation side channels, such as time-varying encoding \cite{Gnanapandithan2025}. Furthermore, bandwidth-limited modulation has been shown to produce source correlations \cite{Yoshino2018,Roberts2018,Lu2021,Kang2023, Pereira2020, Agulleiro2025}. Assuming a decoy state protocol with phase or polarization encoding, these correlations typically fall into one of the following categories -- intensity correlations from decoy state generation, phase correlations from phase randomization, and phase/polarization correlations from phase/polarization encoding.

Multiple studies regarding intensity correlations (commonly referred to as the \textit{pattern effect}) from the use of bandwidth-limited intensity modulators (IM) have been reported thus far \cite{Yoshino2018,Roberts2018,Lu2021,Kang2023, Agulleiro2025}. These studies have a) characterized intensity correlations in GHz-clocked sources and b) proposed/demonstrated various mitigation methods. These methods (based on filtering, precalibration, or improved hardware designs) largely take advantage of the sinusoidal voltage-intensity relationship (specifically, the zero-slope points) of IMs  \cite{Yoshino2018,Roberts2018,Lu2021,Kang2023}. For example, the technique discussed in \cite{Lu2021} reduces correlations by over two orders of magnitude at a 1 GHz repetition rate through the use of dual IMs. Here, a correlation is defined as the percent difference in the intensity of a signal/decoy pulse with respect to the nominal intensity (signal/decoy/vacuum) of the preceding pulse.  

On the other hand, there has been relatively little focus on characterizing and mitigating phase correlations arising from the use of bandwidth-limited PMs. Unlike electro-optic IMs, electro-optic PMs have a linear relationship with voltage, necessitating a separate (albeit analogous) characterization procedure, as well as entirely new mitigation methods. Hence, we report a comprehensive experimental characterization of phase correlations resulting from active (electro-optic) PMs for polarization/phase encoding. We performed this experimental characterization at 500 MHz and 1 GHz, using equipment bandwidths typically used in GHz QKD experiments \cite{Yoshino2018,Lu2021,Kang2023,Roberts2018}. In addition, we performed a simulated characterization at various repetition rates up to 3 GHz.

Fully passive QKD has been proposed as a solution for all modulator side channels at the source, including both intensity and phase/polarization correlations \cite{Wang2023}. Experimental demonstrations have also been reported, albeit at low source repetitions of 20 and 50 MHz \cite{Lu2023}\cite{Hu2023}. While intriguing, the fully passive QKD protocol requires a significant level of post-selection, resulting in an order of magnitude reduction in the secret key rate. In addition, fully passive QKD requires multiple detectors at the source, which limits the source repetition rate and may introduce vulnerabilities due to imperfections in the detectors \cite{Wang2023}.  

Hence, this work highlights an alternative option to fully passive QKD for mitigating source correlations and other side channels arising from PMs -- a \textit{path-selection modulation} QKD source which we demonstrate at a repetition rate of 1 GHz using 1550 nm single-mode fiber, requiring no additional post-selection or the use of detectors. In addition to performing passive polarization encoding using multiple optical paths, it uses gain-switching to generate phase-randomized pulses \cite{Kobayashi2014,Paraso2019}. Although the source does not use phase/polarization modulators, it still uses IMs to select the path taken by each pulse, which determines its polarization state. Specifically, IMs are used only for pulse suppression and transmission in the passive polarization encoding setup. When operated in this on/off manner, any resulting intensity correlations have been shown to be negligible at GHz repetition rates \cite{Yoshino2018,Lu2021}, which we further verify through experimental characterization.

This paper is organized as follows. We start by discussing our experimental and simulated characterization of encoding correlations resulting from PMs in Section \ref{sec:active_characterization}. We then proceed to introduce and discuss our schematic for a path-selection modulation source in Section \ref{sec:schematic}. Lastly, in Section \ref{sec:characterization_sidech}, we experimentally characterize the source. Specifically, we characterize the a) generated polarizations, b) level of phase randomization achieved through gain-switching, c) indistinguishability of the generated polarization states, and d) intensity correlations from using IMs.

\section{Characterization of Correlations from Active Phase Modulation}
\label{sec:active_characterization} 


Here, we present experimental and simulated characterizations of phase correlations (including non-nearest-neighbour correlations) when using electro-optic PMs to perform encoding. Experimental characterization (constrained by equipment bandwidth) was performed at 500 MHz and 1 GHz, while simulated characterization was performed at 500 MHz, 1 GHz, 2 GHz, and 3 GHz. 

\subsection{Experimental Characterization}

The experimental setup shown in Figure \ref{fig:characterization-setup} was used to obtain the characterization results. Phase modulation was achieved by employing an electro-optic IM -- i.e. an electro-optic PM that uses a Mach-Zehnder interferometer to convert a phase change to an intensity change. In this way, modulated phase can be measured and characterized using a photodetector. When the modulator is biased at its minimum, its phase-intensity relationship can be expressed as per Equation \ref{eq:phase-intensity}, where $\phi$ represents phase and $I$ represents normalized intensity. Note that $\phi$ has a linear relationship with the voltage applied to the modulator. 

\begin{equation}
    \phi = 2cos^{-1}(\sqrt{1-I})
    \label{eq:phase-intensity}
\end{equation}

\begin{figure}[h!]
\centering\includegraphics[width=\linewidth]{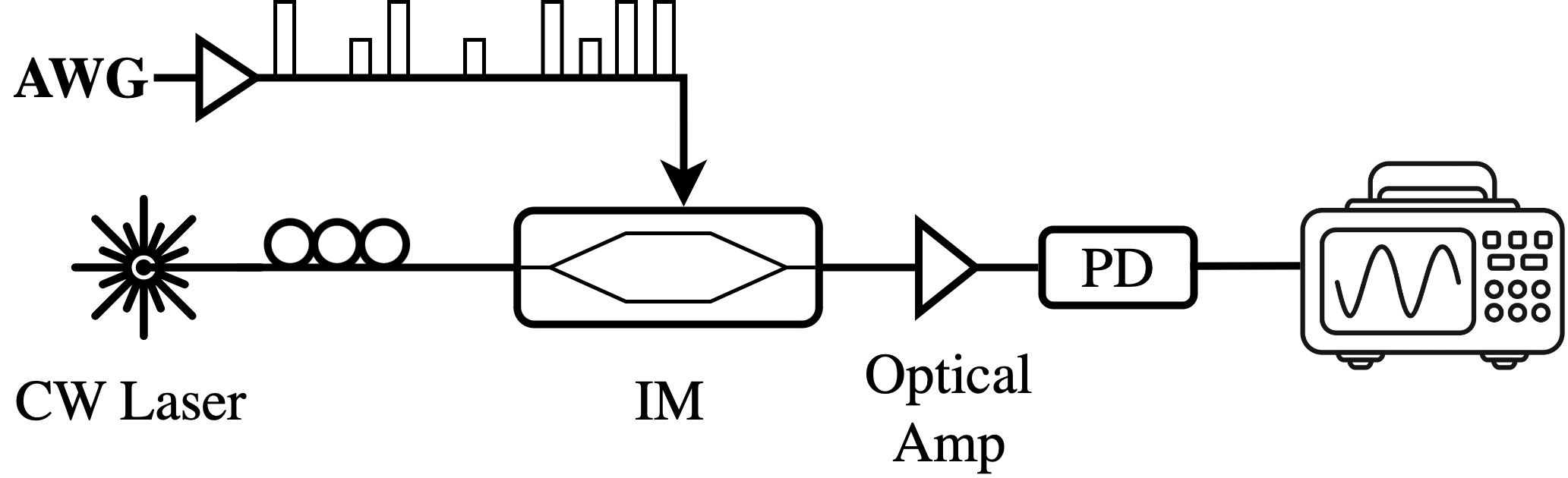}
\caption{Experimental setup for measuring phase correlations at modulation rates of 1 GHz and 500 MHz. The equipment used is as follows. 25 GHz Keysight M8195A arbitrary waveform generator (AWG), 12 GHz ixBlue DR-VE-10-MO RF amplifier, 15 GHz JSDU OC-192 IM, Thorlabs semiconductor optical amplifier, 50 GHz Newport D-8ir photodiode (PD), and 12 GHz Agilent DSO81204A oscilloscope. These bandwidths are typical in GHz QKD experiments \cite{Yoshino2018,Lu2021,Kang2023,Roberts2018}. In this experiment, the IM acts as a PM, converting the phase change into an intensity change using a Mach-Zehnder interferometer.}
\label{fig:characterization-setup}
\end{figure}

To characterize correlations at 1 GHz (500 MHz), the IM was driven to randomly produce one of the three desired phase changes in a three-state protocol, 0, $\pi$/2, and $\pi$, every ns (2 ns). The photodiode and oscilloscope were used to collect the resulting phase/intensity modulated light. For each repetition rate, 15 000 intensity modulation events were collected and processed into phase modulation events (see Section IA of the Supplemental Material for more detailed information on the experimental and data processing methods).  

\begin{figure}[h!]
\centering\includegraphics[width=0.8\linewidth]{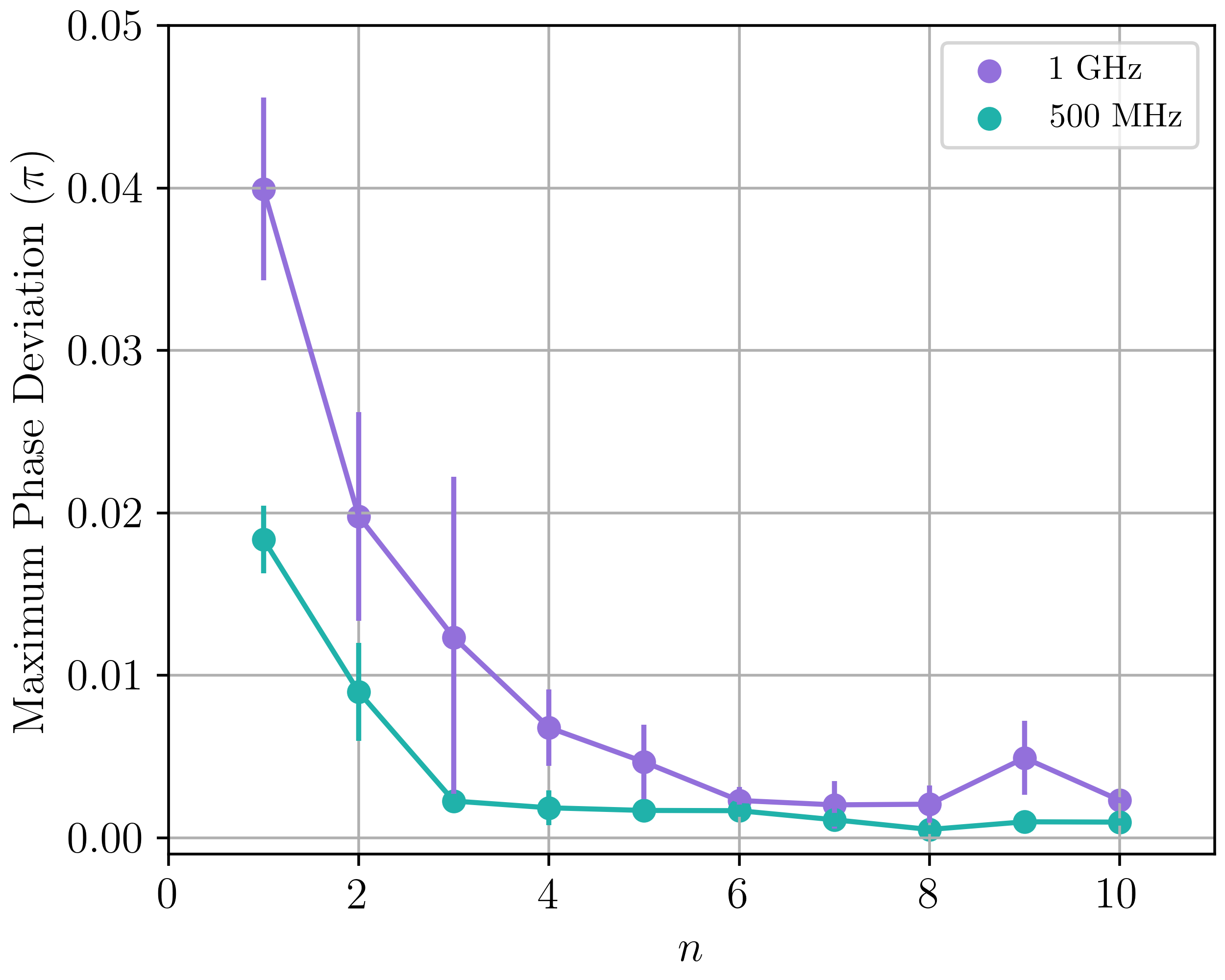}
\caption{Experimental maximum phase deviation of a pulse relative to the nominal phase of the pulse $n$ pulses earlier ($n=1$ denotes the immediately preceding pulse).}
\label{fig:experimental-correlations}
\end{figure}

The goal was to determine how the phase of a pulse differs depending on the nominal phase of the pulse preceding it by $n$ pulses. To denote the directly preceding case, $n=1$ is used. There are 6 cases to consider for each $n$: $0\rightarrow \frac{\pi}{2}$, $\frac{\pi}{2}\rightarrow \frac{\pi}{2}$, $\pi \rightarrow \frac{\pi}{2}$, $0\rightarrow \pi$, $\frac{\pi}{2}\rightarrow \pi$, and $\pi \rightarrow \pi$. All the occurrences of each case were combined into a mean phase and standard deviation. We quantify correlations using the deviation in a mean phase, relative to the case where the preceding pulse has a nominal phase of 0. For example, consider the case where the $0\rightarrow \pi$ case has a mean phase of $\pi$ and the $\pi \rightarrow \pi$ case has a mean phase of $\pi + \epsilon$. The phase deviation for the $\pi \rightarrow \pi$ case would be $\epsilon$. Using the 15 000 phase modulation events, the phase deviation was obtained for each of the 6 cases for each $n$.  To visualize the results (specifically, the magnitude of the correlations and how they change with $n$), the maximum phase deviation among the 6 cases was plotted for each $n$. The results can be seen in Figure \ref{fig:experimental-correlations}. To verify the repeatability of the experiment, two more trials were conducted at each repetition rate -- each time, a different SMA cable was used to connect the AWG and RF amplifier (as this is the longest SMA cable used in the setup). Hence, there are three trials in total for each repetition rate. Figure \ref{fig:experimental-correlations} presents the mean of these three trials, with the error bars representing the standard deviation between the three trials.

The results demonstrate some key points. At 1 GHz, nearest-neighbour correlations are roughly $0.04\pi$, while at 500 MHz, nearest-neighbour correlations are roughly $0.02\pi$.  At 1 GHz, correlations only begin to level out at around $n=6$, while at 500 MHz, correlations begin to level out at around $n=3$. In general, as expected, for a given $n$, the results for $2n$ at 1 GHz roughly correspond to the results for $n$ at 500 MHz. In other words, the correlations for 1 GHz and 500 MHz would be similar if they were plotted against time, rather than $n$. Finally, as expected, correlations increase noticeably as we enter GHz repetition rates -- both the magnitude of the correlations as well as their span across non-nearest-neighbour pulses. 

\subsection{Simulated Characterization}
\label{sec:simulated_characterization}
Here, the phase modulation setup from Figure \ref{fig:characterization-setup} was modelled in order to simulate the phase modulation process. 

A basic Bessel filter was used to model the voltage pulse trains generated by the AWG, using its specified bandwidth of 25 GHz. Low-pass filters such as Bessel filters and Butterworth filters are the most natural choice for modeling bandwidth-limited equipment. A Bessel filter was chosen as it causes the simulator to more closely approximate the experimentally attained waveforms (shown in Section IB of the Supplemental Material in Figure 2).

The RF amplifier was modeled using its magnitude and phase responses, obtained from the datasheet and the manufacturer, while the IM was modeled with its magnitude response and an assumed ideal linear phase response (as its phase response data was unavailable). A basic Bessel filter was used to model the effect of the oscilloscope, using its specified bandwidth of 12 GHz. Additionally, the limited sampling rate (40 GSa/s) of the oscilloscope was incorporated using downsampling. The photodiode, having a 50 GHz bandwidth, was omitted from the simulation as its response far exceeds that of the other components. The primary bandwidth-limiting components of concern are the 12 GHz RF amplifier, 15 GHz IM, and 12 GHz oscilloscope.  

An illustration of the overall simulation process is shown in Section IB of the Supplemental Material in Figure 4. As the phase change applied by the modulator is proportional to the applied voltage, the output of this process can be used to represent the phase-modulated light, with the normalized amplitude being proportional to a phase change (0 corresponding to $0$ and 1 corresponding to $\pi$).

To match the experimental characterization, the same type of driving signal was used for the simulated characterization. In other words, to characterize correlations at 1 GHz (500 MHz), the AWG model was set to randomly produce one of the three desired phase changes in a three-state protocol, 0, $\pi$/2, and $\pi$, every ns (2 ns). And likewise, for other repetition rates. In total, the simulator was used to produce 15 000 phase modulation events for each repetition rate (see Section IB of the Supplemental Material for more detailed information).

\begin{figure}[h!]
\centering\includegraphics[width=\linewidth]{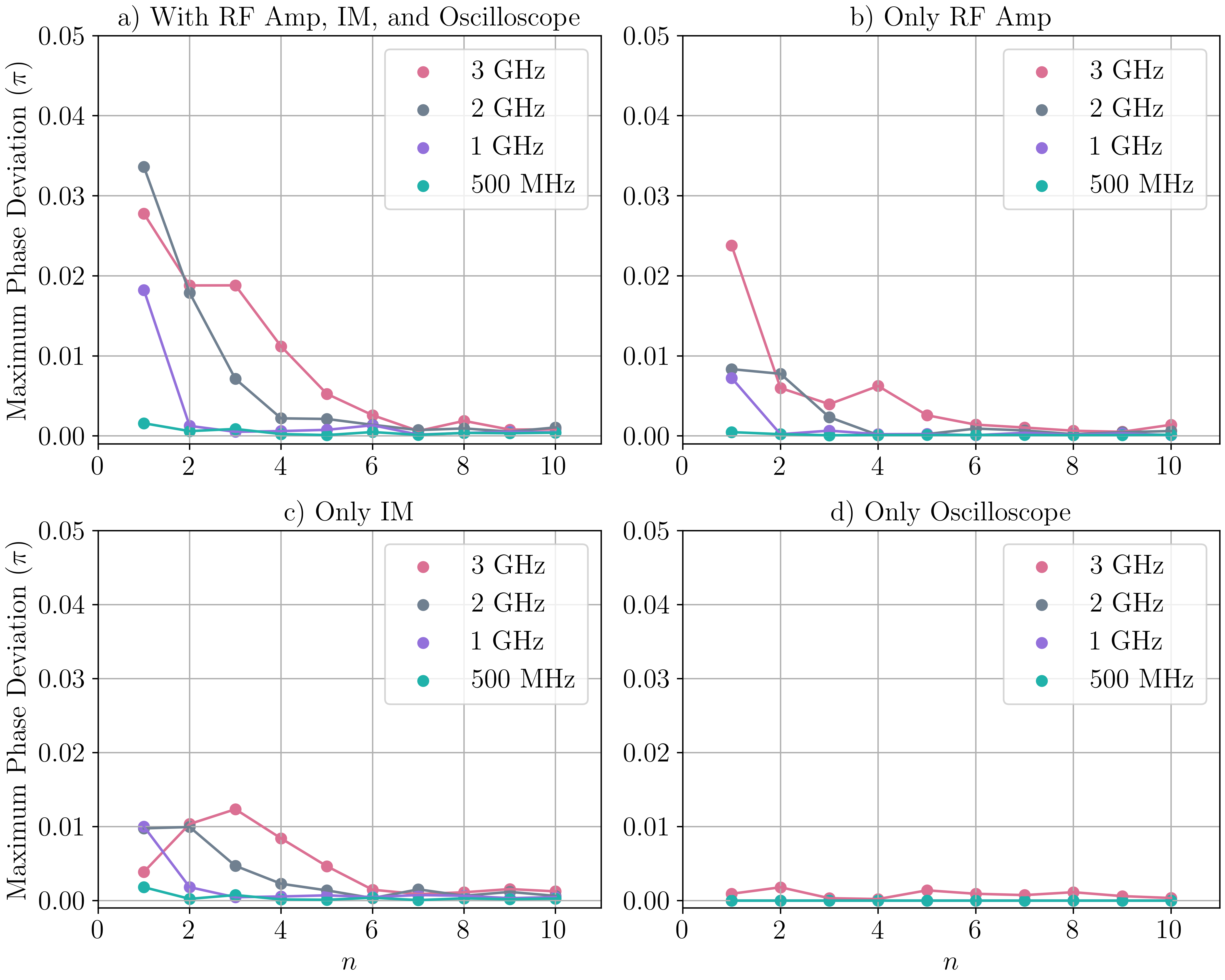}
\caption{Simulated maximum phase deviation of a pulse relative to the nominal phase of the pulse $n$ pulses earlier ($n=1$ denotes the immediately preceding pulse). a) Complete simulation (includes the bandwidth-limiting effects of the RF amplifier, IM, and oscilloscope). b) Simulation with the RF amplifier alone. c) Simulation with the IM alone. d) Simulation with the oscilloscope alone.}
\label{fig:simulated-correlations}
\end{figure}

This data was analyzed in the same manner as the experimental data. For each $n$ and for each of the 6 cases, the phase deviation in the mean phase, relative to the case where the preceding pulse has a nominal phase of 0, was acquired. The corresponding visualization of the correlations (i.e the simulated analogue of Figure \ref{fig:experimental-correlations}) can be seen in Figure \ref{fig:simulated-correlations}a. To provide insight into how the primary bandwidth-limiting components impact the correlations, the results are also shown with the RF amplifier alone (Figure \ref {fig:simulated-correlations}b), the IM alone (Figure \ref{fig:simulated-correlations}c), and the oscilloscope alone (Figure \ref{fig:simulated-correlations}d).  Since repeatability is not a concern for the simulated results, only one trial is shown for each case. 

We first discuss the complete simulation shown in Figure \ref{fig:simulated-correlations}a. Notice that the simulated results at 1 GHz are of the same order of magnitude as the experimental results. Conceptually, this demonstrates that bandwidth limitations are a leading factor contributing to the phase correlations observed in our experiment. Nonetheless, the simulated correlations are noticeably lower than the experimental correlations. There are three main reasons for this. First, idealized approximations were made regarding the unknown equipment responses (namely, the phase response of the IM and the overall response of the oscilloscope). Second, the simulation only captures the linear behavior of the equipment. Hence, the simulation serves more as an optimistic estimate (i.e. lower bound) for the correlations, while the experiment serves more as a pessimistic estimate (i.e. upper bound). Third, the simulation does not account for effects due to cables and connectors.

We now discuss the simulation results shown in Figure \ref{fig:simulated-correlations}bcd, which individually examine the effects of the RF amplifier, IM, and oscilloscope, respectively. The RF amplifier appears to have a noticeable effect on all three GHz repetition rates (1 GHz, 2 GHz, and 3 GHz). The IM appears to have a similar effect, except at 3 GHz, for which nearest-neighbour correlations are noticeably lower. In relation to the RF amplifier and IM, the oscilloscope has a small effect on the correlations.  However, recall that the RF amplifier is the only component for which we know the complete frequency response (magnitude and phase). The frequency response of the oscilloscope was approximated using a generic low-pass filter (in our case, a Bessel filter), while the phase response of the IM was assumed to be linear. Hence, the impact of the IM and oscilloscope is not fully reflected in the simulation. 

Finally, while this work focuses on the experimental/simulated characterization of phase correlations, quantifying the impact of these results on the secret key rate remains an important direction for future work \cite{Curras-Lorenzo2023}. Regardless, the characterization results demonstrate a clear increase in phase correlations with respect to the clock rate due to practical bandwidth limitations in the modulation equipment. Hence, it is vital to explore mitigation strategies if we hope to continue pushing QKD into the GHz clock rate range while maintaining its security advantages. Our path-selection modulation source is one such strategy which we propose and characterize in the remainder of this work.

\section{Design for the Path-Selection Modulation Source}
\label{sec:schematic}
\subsection{High-Level Concept}

\begin{figure}[h!]
\centering\includegraphics[width=\linewidth]{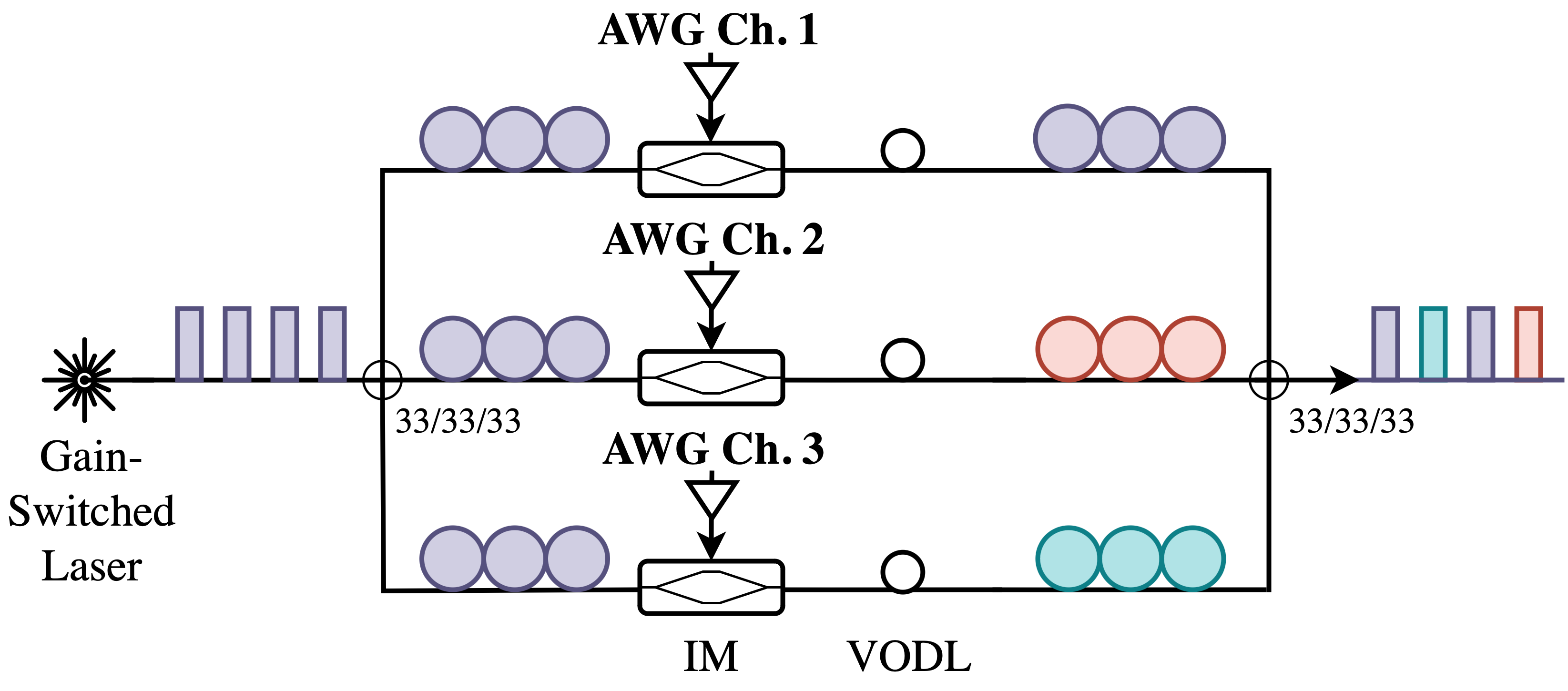}
\caption{Schematic for a 1 GHz path-selection modulation source where each path applies a polarization change corresponding to one of the three polarizations in a three-state protocol. The paths are of equal length. IMs are used to allow a pulse from one random path through every ns. The equipment used is as follows. Eblana Photonics REP1550-DM-H laser diode, Fujitsu FTM7937EZ IM, JDSU OC-192 IM, EOSPACE IM, and Keysight M8195A AWG. Agiltron MDTD variable optical delay lines (VODL) are used to resolve length differences between the three paths.}
\label{fig:source-schematic}
\end{figure}

The design for the path-selection modulation source is shown in Figure \ref{fig:source-schematic}. Evidently, this source does not make use of any active PMs -- hence, there would be no phase/polarization correlations arising from bandwidth-limited phase modulation (active PMs are often used to perform polarization encoding). Phase-randomized pulses are generated through a gain-switched laser (as discussed in Section \ref{sec:gain-switching}) \cite{Kobayashi2014,Paraso2019}. IMs and 1x3 beamsplitters with equal splitting ratio are used to perform encoding, where each path corresponds to one of the three states in a three-state protocol. For concreteness, this work focuses on polarization encoding, although time-bin and differential phase encoding could also be implemented. The IMs are used to ``select'' a path -- i.e. allow a pulse from only one of the paths through, every pulse repetition period. In our experiments (unless stated otherwise), all the IMs are biased at their minimum. Hence, to let a pulse through, an IM would need to be modulated with a voltage pulse of $V_{\pi}$ amplitude (and the voltage pulse would need to be temporally aligned with the corresponding optical pulse). Lastly, optical delay lines  are used to ensure that the length of each path is equal (in our experiment, delay lines of 1 fs resolution were used). The result is a train of polarization-encoded, phase-randomized pulses. The IM operates in only one of two modes -- complete pulse suppression or complete pulse transmission. Both of these modes have been shown to be \textit{stable modes} -- in other words, they are minimally impacted by correlations, as explained in Section \ref{sec:stable-modes} \cite{Yoshino2018,Lu2021}. The IMs and gain-switched laser are all driven by synchronized AWG channels.

\subsection{Stable Modes of an IM}
\label{sec:stable-modes}

Recall from Section \ref{sec:active_characterization} that an electro-optic IM modulates intensity by combining phase modulation with Mach-Zehnder interferometry, resulting in a sinusoidal relationship between voltage and intensity. Notably, at normalized intensities of 0 and 1, the voltage-intensity relationship has a slope of 0. Hence, these points are referred to as \textit{stable points} and are far less susceptible to being affected by the intensity setting of nearby pulses. Some notable works \cite{Yoshino2018,Lu2021} have taken advantage of these points to reduce intensity correlations in decoy-state QKD by over two orders of magnitude. 

As the IMs in the path-selection modulation source serve to either completely transmit or completely suppress optical pulses, they only operate at stable points. To verify that they produce negligible intensity correlations, the intensity correlations that are generated by the source are characterized in Section \ref{sec:intensity_correlations}. 

\subsection{Gain-Switching}
\label{sec:gain-switching}
\textit{Gain-switching} is a useful experimental technique that can be used to generate phase-randomized optical pulses, typically using a semiconductor laser \cite{Kobayashi2014}. In gain-switched semiconductor lasers, the drive current is periodically modulated such that the minimum, $i_{min}$, lies below the threshold current, $i_{threshold}$. Each time the driving current exceeds $i_{threshold}$, an optical pulse is created via a spontaneously emitted (with random phase) seed photon. 

However, there is concern as to whether the phase-randomized property of these pulses persists at high (GHz) repetition rates. As reported in \cite{Kobayashi2014}, at GHz repetition rates, a photon from a pulse may survive long enough to act as a seed for a subsequent pulse. This would result in phase correlations between the pulses. \cite{Kobayashi2014} suggests that this problem can be mitigated by setting $i_{min}$ sufficiently below $i_{threshold}$. To verify this for our path-selection modulation source, we characterize the level of phase randomization for various $i_{min}$ in Section \ref{sec:phase_rand}.

\section{Characterization of the Path-Selection Modulation Source}
\label{sec:characterization_sidech}

\subsection{Polarization Drift}
Here, the three polarization states produced by the source are characterized. To characterize the polarization produced by a specific path $x$, the IM on that path was configured to transmit all pulses, while the IMs on all other paths were set to suppress their pulses. As a result, the output of the source consisted solely of pulses originating from path $x$. This output was connected to a polarimeter and the polarization was collected over 2 hours, once a minute.

\begin{figure}[h!] 
    \centering
    \includegraphics[width=\columnwidth]{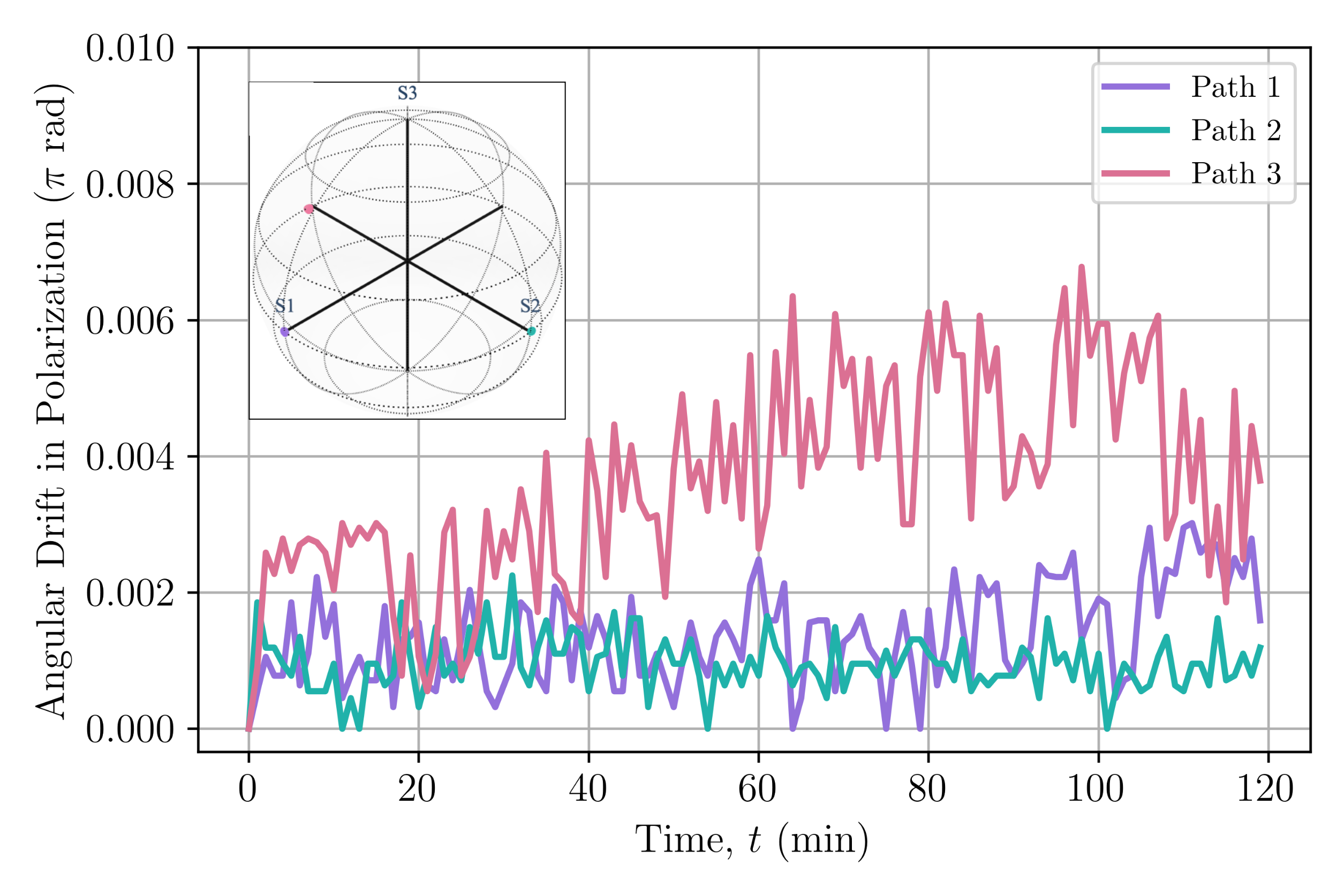}
    \caption{Angular distance between the polarization at time $t$ and the polarization at time 0 for each path of the path-selection modulation source. The inset illustrates the polarization trace over 2 hours for each path on a Poincare sphere.}
    \label{fig:drift}
\end{figure}

For each path, the change in polarization (with respect to the polarization at time $t=0$) was plotted, as shown in Figure \ref{fig:drift}. Here, the change in polarization at time $t$ is given as the angular distance between the polarization at time $t$ and the polarization at time $0$. Overall, in a lab environment, we see that the polarization drift never exceeded 0.007$\pi$ with respect to this angular distance.

\subsection{Phase Randomization}
\label{sec:phase_rand}
Here, we experimentally characterize the level of phase randomization in the pulses generated by the gain-switched laser for various minimum driving currents, $i_{min}$. As the gain-switched laser is driven by both an AC sine wave source and DC power supply, both the maximum and minimum of the driving current can be controlled.

\begin{figure}[h!] 
    \centering
    \includegraphics[width=\columnwidth]{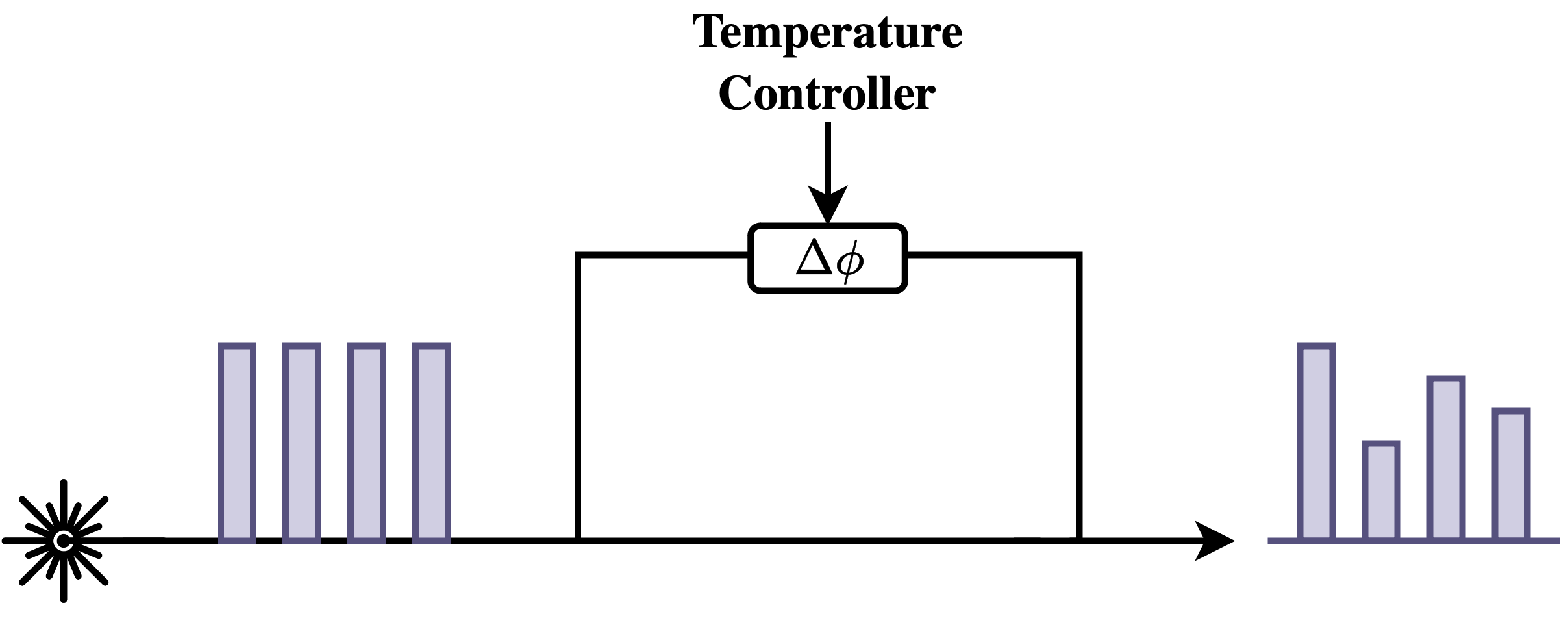}
    \caption{Asymmetric Mach-Zehnder interferometer setup used to characterize the phase randomness of the pulses generated by the gain-switched laser in the  path-selection modulation source. The path length difference between the two arms of the asymmetric interferometer is specified to be 500 ps. Hence, to characterize phase correlations between adjacent pulses, the gain-switched laser was driven at 2 GHz. One arm of the interferometer contains a phase modulator with a thermo-electric cooler (TEC), whose temperature determines the phase change applied -- hence, a TEC controller (i.e. temperature controller) was used to control/stabilize the phase change along this arm.}
    \label{fig:interferometry setup}
\end{figure}

The asymmetric interferometer setup presented in Figure \ref{fig:interferometry setup} was used to perform this characterization. The specific asymmetric interferometer that was employed for this experiment has a path length difference of 500 ps. Hence, the interferometer can interfere adjacent pulses at a repetition rate of 2 GHz. The interference data can be used to quantify the level of phase randomization between adjacent pulses. The specific methods are as follows.

The output of the setup in Figure \ref{fig:interferometry setup} was connected to a slow ($\sim$ seconds) photodiode and power meter. $\Delta \phi$ was varied and the corresponding intensity reading on the power meter was acquired. Using this data (intensity with respect to $\Delta \phi$), the interference fringe visibility, $\mathcal{V}$ (expressed below in Equation \ref{eq:visibility}) can be expressed using the maximum intensity, $I_{max}$ and the minimum intensity, $I_{min}$. Here, $\mathcal{V}$  is the parameter which quantifies the level of phase randomization between adjacent pulses (a high $\mathcal{V}$ would indicate a low level of phase randomization and vice versa). 

\begin{equation}
    \mathcal{V}=\frac{I_{max}-I_{min}}{I_{max}+I_{min}}
    \label{eq:visibility}
\end{equation}

\begin{figure}[h!] 
    \centering
    \includegraphics[width=\linewidth]{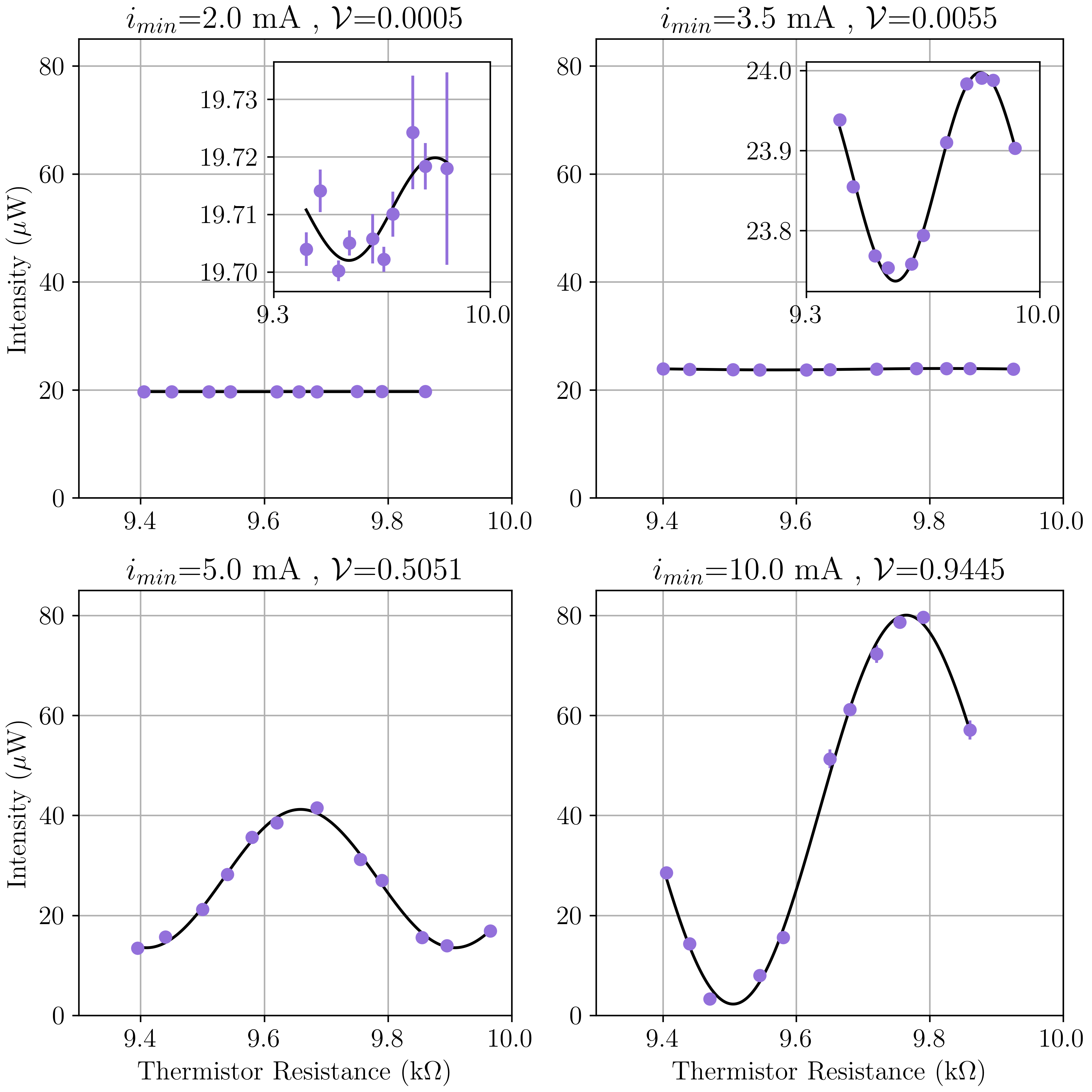}
    \caption{Interference fringes when the gain-switched laser was driven at various minimum driving currents. The thermistor resistance is proportional to the temperature of the TEC, which is proportional to $\Delta \phi$. It is the quantity that is displayed by the TEC controller to express the temperature of the TEC.}
    \label{fig:fringes}
\end{figure}

Intensity data with respect to $\Delta \phi$ (60 intensity readings over a minute for each $\Delta \phi$) was collected for the following $i_{min}$ -- 10 mA, 5 mA, 3.5 mA, and 2 mA (a range of values below the threshold current, 12 mA). The maximum driving current was kept constant at 25 mA. As per the findings in \cite{Kobayashi2014}, $\mathcal{V}$ should decrease (i.e. phase randomization should improve) as $i_{min}$ decreases with respect to the threshold current. The results are plotted in Figure \ref{fig:fringes}, where each point is the mean of the 60 intensity readings and the corresponding error bar is the standard deviation of those 60 readings. As expected, the visibility of the fringes decreases noticeably with the minimum driving current. Notably, at $i_{min}=$ 2mA, $\mathcal{V}$ is on the order of $10^{-4}$. This is well within (by an order of magnitude) the target visibilities specified in \cite{Kobayashi2014} for minimizing signal/decoy distinguishability, as well as basis indistinguishability.

\subsection{State Indistinguishability}
\label{sec:indistinguishability}
In the path-selection modulation source, each path corresponds to a distinct state in a three-state protocol. Hence, any differences in these three paths can be exploited by an eavesdropper to gain information on the states sent. Fortunately, all the pulses, regardless of their state, originate from the same laser and are in the same spatial mode of the single-mode fiber. This leaves two key factors -- intensity and timing. Specifically, the intensity of the pulses from the three paths must be the same and the three paths must have the same path length.

To characterize the indistinguishability of the three paths in intensity and time, the path-selection modulation source was driven at 1 GHz. Here, the IMs were modulated such that the three paths were transmitted in uniformly random order by the source (the overall pulse train was 100 ns long). 

\begin{figure}[h!] 
    \centering
    \includegraphics[width=\linewidth]{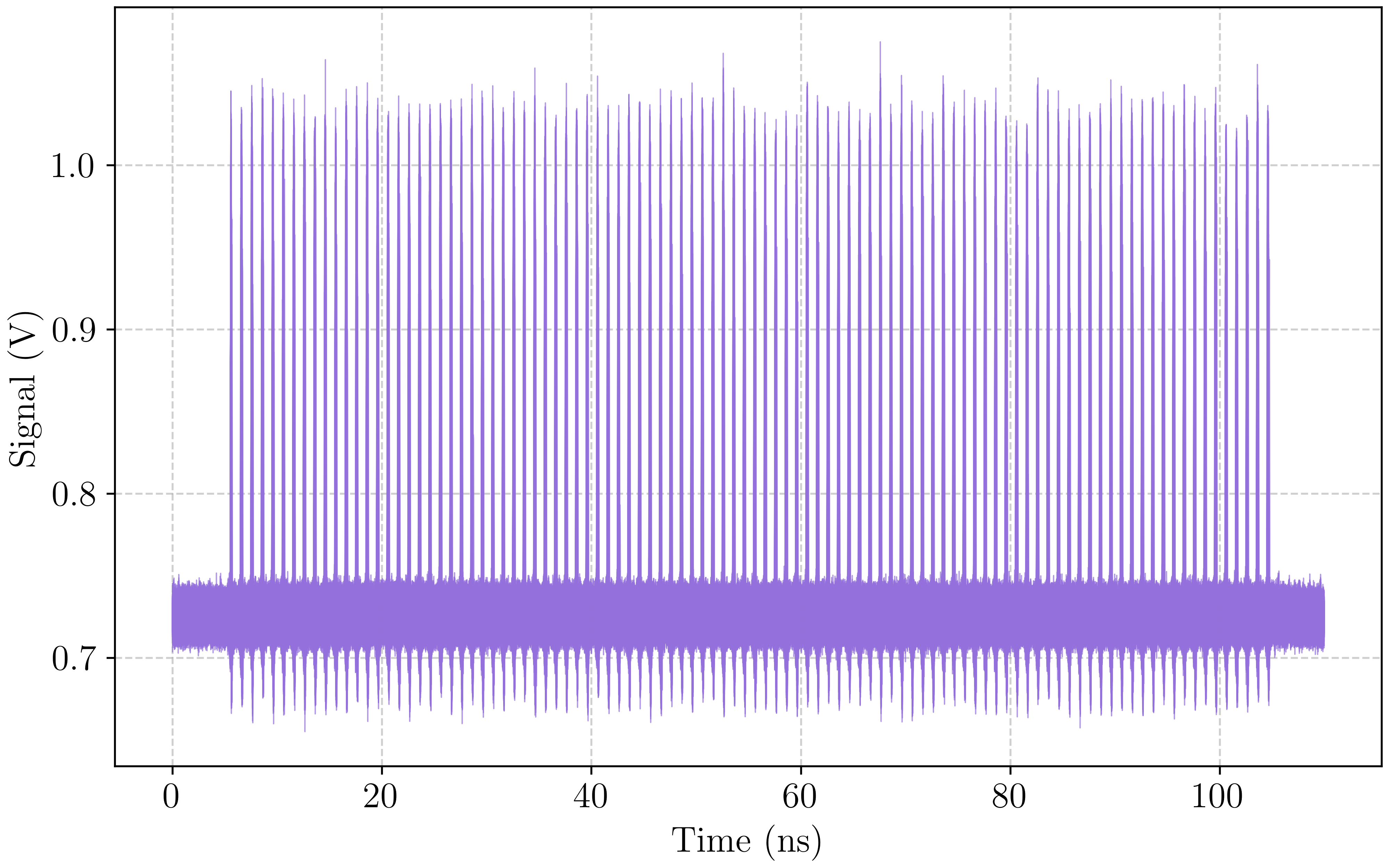}
    \caption{The output pulse train of the path-selection modulation source in persistence mode when the three paths are transmitted in uniformly random order. This plot consists of 4k oscilloscope traces.}
    \label{fig:persistence}
\end{figure}

The output of the source was connected to the 12 GHz Agilent DSO81204A oscilloscope through the Newport AD-10ir amplified photodiode. The 100 pulse long pattern was collected 4000 times and plotted in persistence mode, as shown in Figure \ref{fig:persistence}. 

\begin{figure}[h!] 
    \centering
    \includegraphics[width=\linewidth]{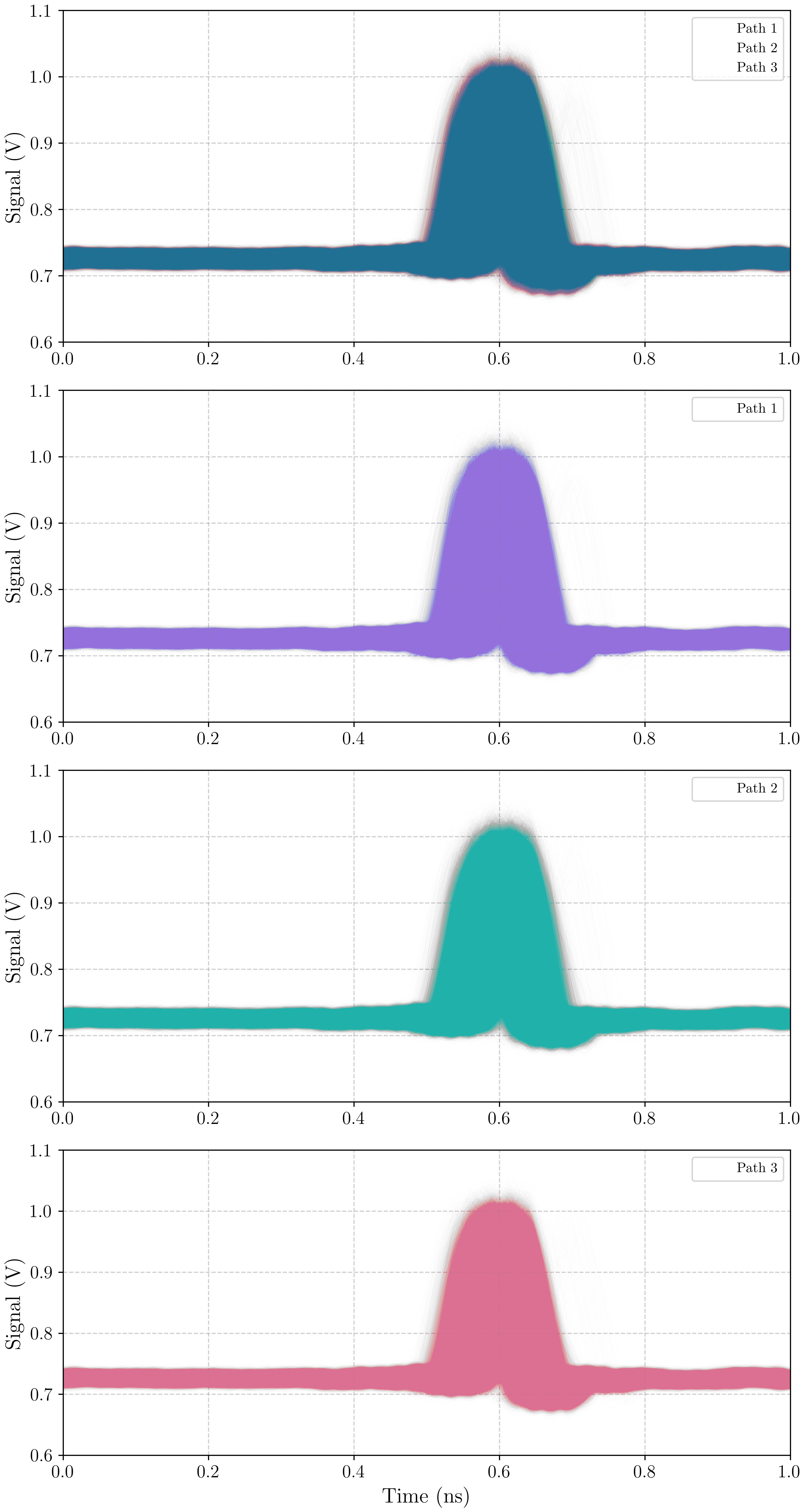}
    \caption{Persistence traces of the pulses produced by each of the three paths in the path-selection modulation source. The topmost plot overlays the three persistence plots for direct comparison. The persistence trace plot corresponding to each path consists of 100k oscilloscope traces.}
    \label{fig:persistence_pulse}
\end{figure}

For each path, the peak pulse intensities and locations were collected from all the acquired oscilloscope traces. The results can be seen in Tables \ref{table:intensity} and \ref{table:time}. Furthermore, for each path, all the collected pulse traces were plotted in persistence, as shown in the bottom three plots in Figure \ref{fig:persistence_pulse}. In the topmost plot, the three persistence plots are shown together, overlaid for direct comparison.

Notably, Table \ref{table:intensity} shows that the differences in peak pulse intensity are two orders of magnitude smaller than the intensities themselves and fall well within their standard deviations. Similarly, Table \ref{table:time} shows that the differences in pulse peak timing are two orders of magnitude smaller than the pulse width (approximately 40 ps) and fall well within the timing standard deviations. Figure \ref{fig:persistence_pulse} further demonstrates the low level of distinguishability between the three paths by comparing their corresponding pulse shape. This distinguishability was also quantified for each pair of paths using a normalized inner product. Specifically, to quantify the distinguishability between path $i$ and path $j$, a representative pulse trace for each path was obtained by averaging its individual oscilloscope traces. The distinguishability between the two averaged pulse traces was then measured using $\epsilon=1-\text{(normalized inner product)}$. The results can be seen in Table \ref{table:inner_prod1}.

Note that our experiment does not employ identical IMs across the three paths, as illustrated in Figure \ref{fig:source-schematic}. Nonetheless, we were able to achieve this degree of path indistinguishability. In practice, however, using identical IMs for all three paths would further enhance this indistinguishability. 

\begin{table}[h]
    \centering
    \begin{tabular}{||c|c|c||}
        \hline
        Path & Mean (mV) & Standard Deviation (mV) \\ \hline\hline
        1 & 231.6 & 29 \\ \hline
        2 & 233.6 & 28 \\ \hline
        3 & 230.3 & 33 \\ \hline
    \end{tabular}
    \caption{Pulse peak intensity (mean and standard deviation) corresponding to each path of the source, neglecting the DC offset imposed by the photodiode.}
    \label{table:intensity}
\end{table}

\begin{table}[h]
    \centering
    \begin{tabular}{||c|c|c||}
        \hline
        Path & Mean (ps) & Standard Deviation (ps) \\ \hline\hline
        1 & 585.4 & 22 \\ \hline
        2 & 586.2 & 22 \\ \hline
        3 & 586.0 & 22 \\ \hline
    \end{tabular}
    \caption{Pulse peak timing relative to a perfect 1 GHz signal (mean and standard deviation) corresponding to each path of the source.}
    \label{table:time}
\end{table}

\begin{table}[h]
    \centering
    \begin{tabular}{||c|c||}
        \hline
        Paths & Pulse Distinguishability ($\epsilon$) \\ \hline\hline
        1,2 & $5.330 \times 10^{-6}$  \\ \hline
        1,3 & $2.396 \times 10^{-7}$  \\ \hline
        2,3 & $5.834 \times 10^{-6}$  \\ \hline
    \end{tabular}
    \caption{Distinguishability in averaged pulse trace, quantified using $\epsilon=1-\text{(normalized inner product)}$.}
    \label{table:inner_prod1}
\end{table}

\subsection{Intensity Correlations}
\label{sec:intensity_correlations}
As IMs are used in the path-selection modulation source, there is a concern as to whether they would introduce intensity correlations. However, as discussed in Section \ref{sec:stable-modes}, the IMs are only operated at their stable point -- hence, intensity correlations should be minimized. Here, experimental results are presented to verify this.

To proceed, we must first delineate the nature of the intensity correlations that are potentially created by the path-selection modulation source. Consider the voltage pulse train that is used to drive each IM. The IMs serve to transmit a pulse from only one of the three paths in the source (each path representing a polarization from a three-state protocol) every ns (at a GHz clock rate). Meanwhile, the other two paths are suppressed. In other words, a given IM would either transmit an optical pulse (using a voltage pulse of $V_{\pi}$ amplitude) with 1/3 probability or completely suppress an optical pulse (using a voltage pulse of 0 amplitude -- i.e. no pulse) with 2/3 probability. Hence, each IM would receive a constant $V_{\pi}$ amplitude pulse train with the distance between directly adjacent pulses being a random multiple of 1 ns. At GHz repetition rates, the amplitude of a pulse could vary based on the amplitude of a preceding pulse -- by extension, the amplitude of a pulse may also vary based on the location of a preceding pulse. Hence, the path-selection modulation source may generate intensity-correlated pulses -- except here, the intensity deviation of a pulse is dependent on the location of the preceding pulse with the same polarization. 

\begin{figure}[h!] 
    \centering
    \includegraphics[width=\linewidth]{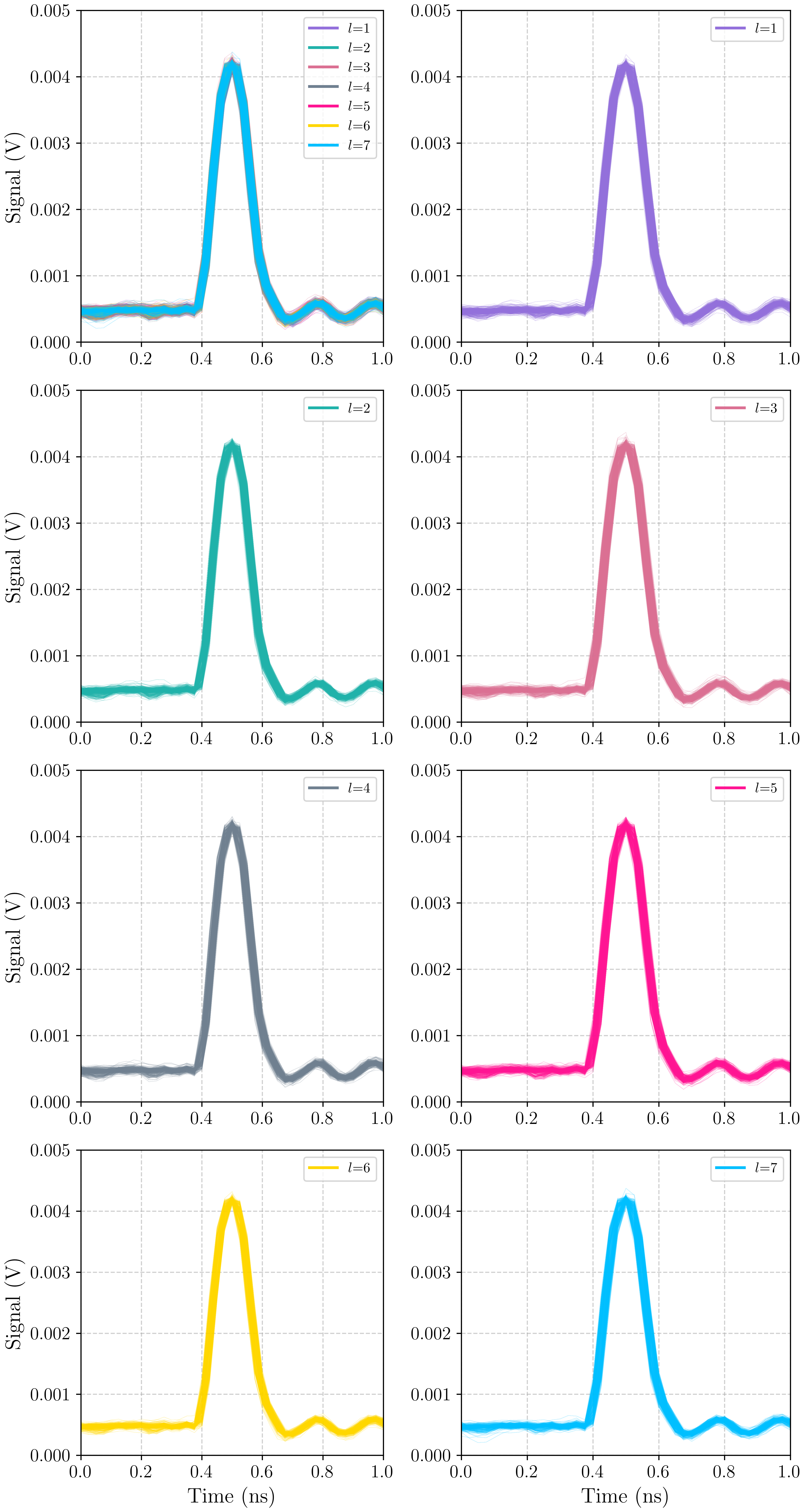}
    \caption{Time-averaged oscilloscope traces (plotted in persistence) for various $l$. The top-left most plot overlays the plots for all $l$ for direct comparison. The persistence trace plot corresponding to each $l$ consists of approximately 300 time-averaged oscilloscope traces.}
    \label{fig:intensity_correlations}
\end{figure}

Hence, the goal of this experiment was to characterize this type of intensity correlation at a repetition rate of 1 GHz. The same experimental setup from Figure \ref{fig:characterization-setup} was used for this. Voltage pulse trains with an amplitude of $V_{\pi}$ and random multiples of 1 ns between adjacent pulses were used to drive the IM. The resulting intensity modulated light was collected using the photodiode and oscilloscope. Overall, 10 000 ns of (time-averaged, to mitigate random noise) data was collected (see Section II in the Supplemental Material for more detailed information on the experimental methods). From this data, the goal was to determine how the intensity of a pulse differs depending on how far away (in ns) the preceding pulse is (here, this distance is denoted as $l$). For each $l$ between 1 ns and 7 ns, all occurrences were combined into a mean intensity and standard deviation. The tabulated results can be found in Table \ref{table:passive-correlations}. To demonstrate the low level of distinguishability in pulse shape, for each $l$ between 1 ns and 7 ns, all corresponding time-averaged pulses were plotted in persistence in Figure \ref{fig:intensity_correlations}. In the top-left most plot, all seven persistence plots are shown together, overlaid for direct comparison. 

\begin{table}
\begin{center} 
\begin{tabular}{||c c c c||} 
\hline
$l$ (ns) & Mean Intensity  & Mean Normalized  & Num  \\
 & (mV) & Intensity &  \\
\hline\hline 
1 & 2.65 $\pm$ 0.04 & 0.996 $\pm$ 0.021 & 354 \\ \hline
2 & 2.65 $\pm$ 0.04 & 0.996 $\pm$ 0.021 & 365 \\ \hline
3 & 2.66 $\pm$ 0.04 & 1.000 $\pm$ 0.021 & 343 \\ \hline
4 & 2.65 $\pm$ 0.04 & 0.996 $\pm$ 0.021 & 296 \\ \hline
5 & 2.65 $\pm$ 0.04 & 0.996 $\pm$ 0.021 & 308 \\ \hline
6 & 2.66 $\pm$ 0.04 & 1.000 $\pm$ 0.021 & 280 \\ \hline
7 & 2.66 $\pm$ 0.04 & 1.000 $\pm$ 0.021 & 285 \\
\hline 
\end{tabular}
\end{center}
\caption{Mean intensity of optical pulses with varying distances from the preceding pulse. The standard deviation is given here as the uncertainty in the mean intensity. The mean normalized intensity is the mean intensity divided by the maximum mean intensity. The number of intensities used to compute each mean is also shown.} 
\label{table:passive-correlations}
\end{table}

\begin{table}[h]
    \centering
    \begin{tabular}{||c|c||}
        \hline
        $l$ (ns) & Maximum Pulse Distinguishability ($\epsilon$) \\ \hline\hline
        1 & $1.231 \times 10^{-4}$  \\ \hline
        2 & $4.441 \times 10^{-5}$  \\ \hline
        3 & $1.444 \times 10^{-4}$  \\ \hline
        4 & $1.444 \times 10^{-4}$  \\ \hline
        5 & $1.156 \times 10^{-4}$  \\ \hline
        6 & $4.869 \times 10^{-5}$  \\ \hline
        7 & $5.748 \times 10^{-5}$  \\ \hline
    \end{tabular}
    \caption{Maximum (with respect to other $l$ values) distinguishability in averaged pulse trace for each $l$, quantified using $\epsilon=1-\text{(normalized inner product)}$.}
    \label{table:inner_prod2}
\end{table}

As expected, there is no noticeable correlation between $l$ and the intensity of a pulse. Furthermore, any deviation in the intensity is on the order of 0.1\% (consistent with prior GHz characterizations of intensity correlations at stable points \cite{Yoshino2018,Lu2021}) and well within the standard deviation. Figure \ref{fig:intensity_correlations} further demonstrates the level of indistinguishability by comparing the pulse shape for every $l$. The distinguishability was also quantified for each pair of $l$ values using the $\epsilon$ parameter described in Section \ref{sec:indistinguishability}. In Table \ref{table:inner_prod2}, the maximum $\epsilon$ value is given for each $l$.

\section{Conclusion}
\label{sec:conclusion}

Typically, due to bandwidth limitations, active modulators can introduce correlations at GHz clock rates \cite{Yoshino2018,Roberts2018,Lu2021,Kang2023}. In contrast to previous work which focuses on correlations due to IMs, here we focus on correlations due to PMs. Specifically, we have reported an experimental and simulated characterization of phase correlations resulting from active (electro-optic) phase modulation for polarization/phase encoding. We performed our characterization at various repetition rates using equipment bandwidths typically used in GHz QKD experiments \cite{Yoshino2018,Lu2021,Kang2023,Roberts2018}. Our results suggest that at GHz repetition rates, nearest-neighbour correlations are on the order of $0.01\pi$.

In general, active modulators introduce a myriad of security vulnerabilities, of which correlations are only one example \cite{Gnanapandithan2025,Bourassa2022}. As a potential solution to this issue, we have proposed a path-selection modulation source and characterized it at a clock rate of 1 GHz, using polarization encoding. As this source does not make use of any active phase/polarization modulation components, it does not introduce phase/polarization correlations. Although the source makes use of IMs for path-selection modulation, they introduce negligible intensity correlations as they are only operated at their stable points. We experimentally confirm this by quantifying the intensity deviations, demonstrating that they remain within approximately 0.1\%. Furthermore, rather than using active PMs, our source uses gain-switching to generate phase-randomized optical pulses. We use interference fringe visibility to quantify the level of phase randomization, achieving a visibility on the order of $10^{-4}$. This is well within (by an order of magnitude) the target visibilities specified in \cite{Kobayashi2014} for minimizing signal/decoy distinguishability, as well as basis indistinguishability.

Moreover, as the generated polarization states originate from different paths, we demonstrated indistinguishability in the following degrees of freedom -- optical intensity and pulse timing. We demonstrate that the differences in peak pulse intensity are two orders of magnitude smaller than the intensities themselves and fall well within their standard deviations. We demonstrate that the differences in pulse peak timing are two orders of magnitude smaller than the pulse width (approximately 40 ps) and fall well within the timing standard deviations. We also demonstrate that polarization drift in the generated states stays below $0.007\pi$ over two hours in a lab environment.

We remark that fully passive QKD has an advantage in that it eliminates all modulator side channels, while our source only mitigates side channels arising from PMs. Although it is possible to strategically employ active IMs to mitigate intensity correlations \cite{Yoshino2018,Lu2021,Kang2023}, other side channels created by IMs continue to pose a risk. One such vulnerability is the temporal dimension of pulses which may leak information on whether a pulse is a signal or decoy. An analogous issue concerning temporal pulse phase profiles has recently been reported in \cite{Gnanapandithan2025} and shown to possibly lower the secret key rate by several orders of magnitude. On the other hand, fully passive QKD is subject to security risks and limitations on repetition rate due to its use of imperfect practical detectors at the source. Furthermore, fully passive QKD requires a significant level of post-selection, which results in an order of magnitude reduction of the secret key rate. None of these factors are of concern for our path-selection modulation source.

\begin{acknowledgments}
We would like to thank Andi Shahaj, Eric Min, and Abdullah Fawzy for their generous help with the experimental setups. We would also like to thank Wenyuan Wang for valuable discussions and inspiration during the early conceptual stage of this work. This research was supported in part by NSERC (Alliance Grant), CFI, ORF, MITACS, DRDC, and National University of Singapore (start-up grant).
\end{acknowledgments}

\bibliographystyle{apsrev}
\bibliography{biblio}

\widetext
\clearpage

\begin{center}
    \textbf{Supplemental Material}
\end{center}

\setcounter{equation}{0}
\setcounter{figure}{0}
\setcounter{table}{0}
\setcounter{section}{0}
\setcounter{page}{1}

\makeatletter
\renewcommand{\theequation}{S\arabic{equation}}
\renewcommand{\thefigure}{S\arabic{figure}}

\section{Details Regarding the Characterization of Correlations Using Active Modulation}
\subsection{Experimental Methods}
\label{appendix:correlations_characterization_exp_methods}

See Figure \ref{fig:characterization-setup} for the experimental setup, reproduced from the main text. 150 voltage pulse trains, each with a repetition rate of 1 GHz and a pulse width of 200 ps (a modulation width commonly used in GHz QKD experiments \cite{Lu2021,Kang2023,Roberts2018,Yoshino2018,Li2019}) were generated by the arbitrary waveform generator (AWG), amplified by the RF amplifier, and used to drive the intensity modulator (IM). RF amplification was necessary, as a modulation amplitude of 5-6 V is typically required by IMs to perform a phase change of $\pi$ (the Keysight M8195A high-speed AWG used in this experiment cannot generate amplitudes above 1 V). Each pulse train was pseudorandom, consisting of a repeating set of 100 pulses with randomized amplitude. These amplitudes correspond to one of the three desired phase changes in a three-state protocol (0, $\pi$/2, and $\pi$).

\begin{figure}[h!]
\centering\includegraphics[width=0.4\linewidth]{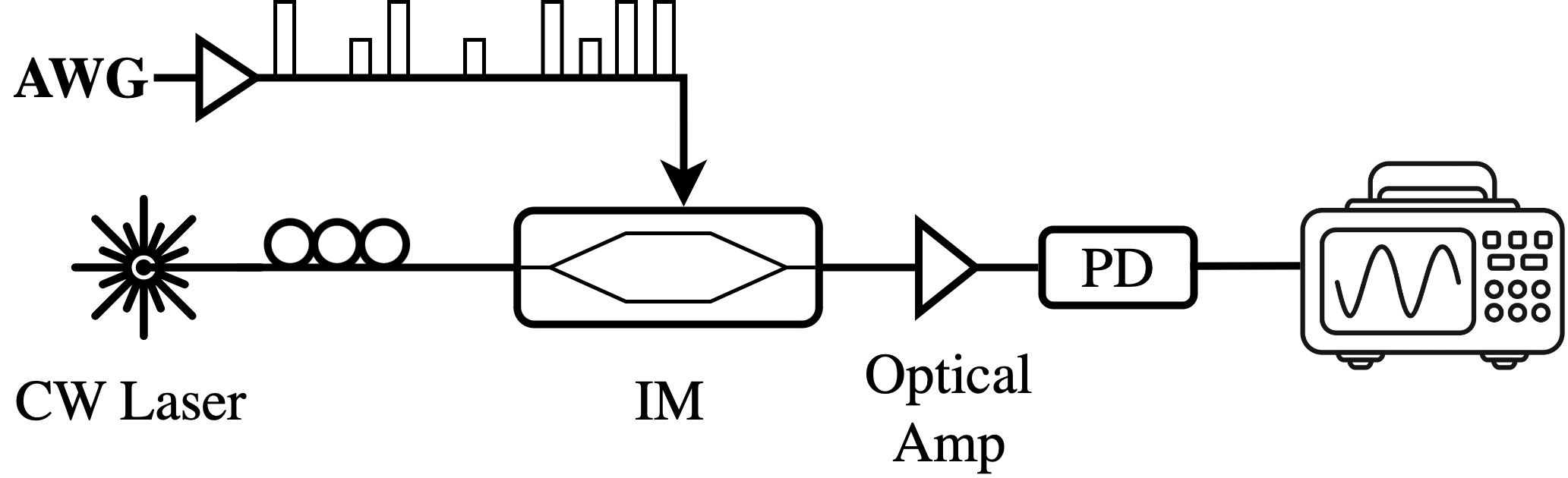}
\caption{Experimental setup for measuring phase correlations at modulation rates of 1 GHz and 500 MHz. The equipment used is as follows. 25 GHz Keysight M8195A AWG, 12 GHz ixBlue DR-VE-10-MO RF amplifier, 15 GHz JSDU OC-192 IM, Thorlabs semiconductor optical amplifier, 50 GHz Newport D-8ir photodiode (PD), and 12 GHz Agilent DSO81204A oscilloscope. These bandwidths are typical in GHz QKD experiments \cite{Yoshino2018,Lu2021,Kang2023,Roberts2018}. In this experiment, the IM acts as a phase modulator, converting the phase change into an intensity change using a Mach-Zehnder interferometer.}
\label{fig:characterization-setup}
\end{figure}

For each of the 150 pulse trains, the corresponding phase (i.e. intensity) modulated light was measured using a photodiode and oscilloscope, after being amplified using an optical amplifier (optical amplification was necessary to improve the signal-to-noise ratio). The entire 100 pulse pattern (corresponding to 100 ns at 1 GHz) was obtained each time. As the pulse trains consisted of repeating sets, the averaging mode could be used on the oscilloscope to average out the effect of short-term noise (an averaging mode of 1024 was observed to be sufficient). 

The same procedure was repeated at 500 MHz. In summary, 150 averaged oscilloscope waveforms (each 100 ns wide, corresponding to a pattern length of 100) were collected at 1 GHz. 150 averaged oscilloscope waveforms (each 200 ns wide, corresponding to a pattern length of 100) were collected at 500 MHz. In total, 15 000 phase (i.e. intensity) modulated pulses were obtained at each repetition rate. 

\begin{figure}[h!]
\centering\includegraphics[width=\linewidth]{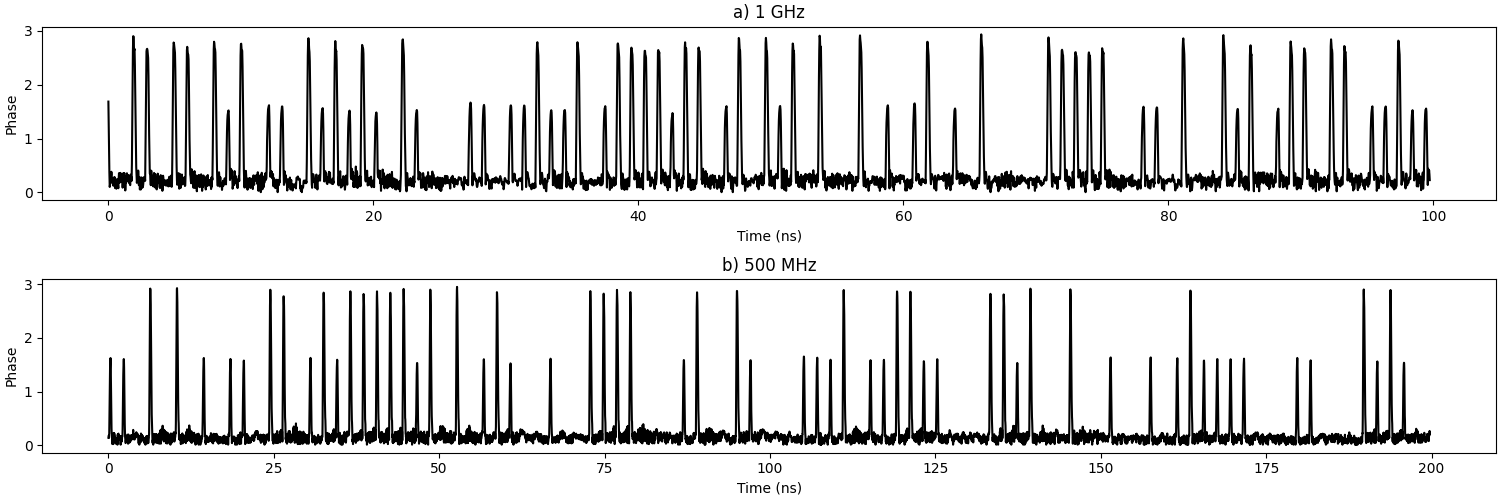}
\caption{Examples of optical phase waveforms at a) 1 GHz and b) 500 MHz. These waveforms were obtained from the raw oscilloscope data (see Figure \ref{fig:voltage-waveform-examples}) by first converting them to normalized intensity waveforms (by removing the photodiode offset and normalizing using the maximum voltage from all the collected waveforms at a given repetition rate). Then, Equation \ref{eq:phase-intensity} was used to convert them to optical phase waveforms.}
\label{fig:phase-waveform-examples}
\end{figure}

\begin{figure}[h!]
\centering\includegraphics[width=\linewidth]{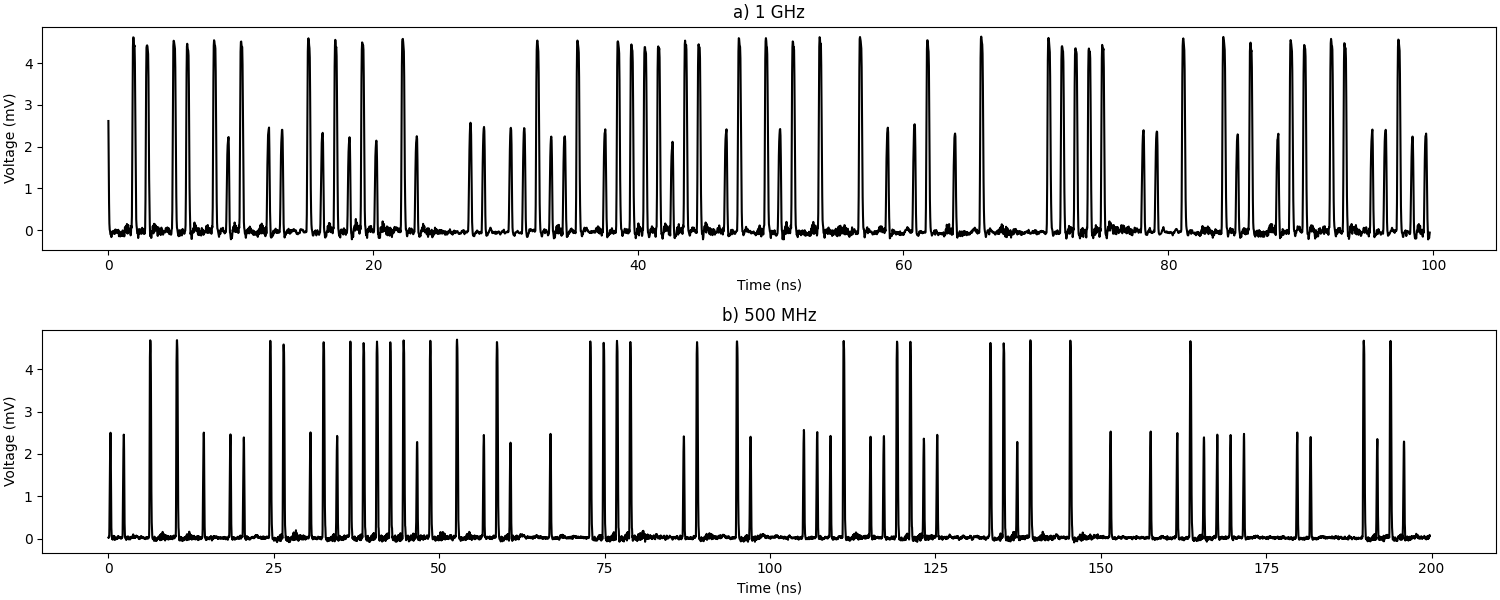}
\caption{Examples of raw voltage waveforms at a) 1 GHz and b) 500 MHz.}
\label{fig:voltage-waveform-examples}
\end{figure}

The Newport D-8ir photodiode introduces a DC offset which was subtracted from the collected oscilloscope waveforms. The resulting waveforms were assumed to be proportional to optical intensity, as the photodiode was operated within its linear response region. For a given repetition rate, the oscilloscope waveforms were normalized using the maximum voltage from all the collected oscilloscope waveforms at that repetition rate. This converts the oscilloscope waveforms into normalized optical intensity waveforms. Then, Equation \ref{eq:phase-intensity} was used to convert the normalized optical intensity waveforms into optical phase waveforms. Examples of these optical phase waveforms are shown in Figure \ref{fig:phase-waveform-examples}. The corresponding raw oscilloscope waveforms are shown in Figure \ref{fig:voltage-waveform-examples}.

\begin{equation}
    \phi = 2cos^{-1}(\sqrt{1-I})
    \label{eq:phase-intensity}
\end{equation}

\subsection{Simulation Methods}
\label{appendix:correlations_characterization_sim_methods}

\begin{figure}
\centering\includegraphics[width=\linewidth]{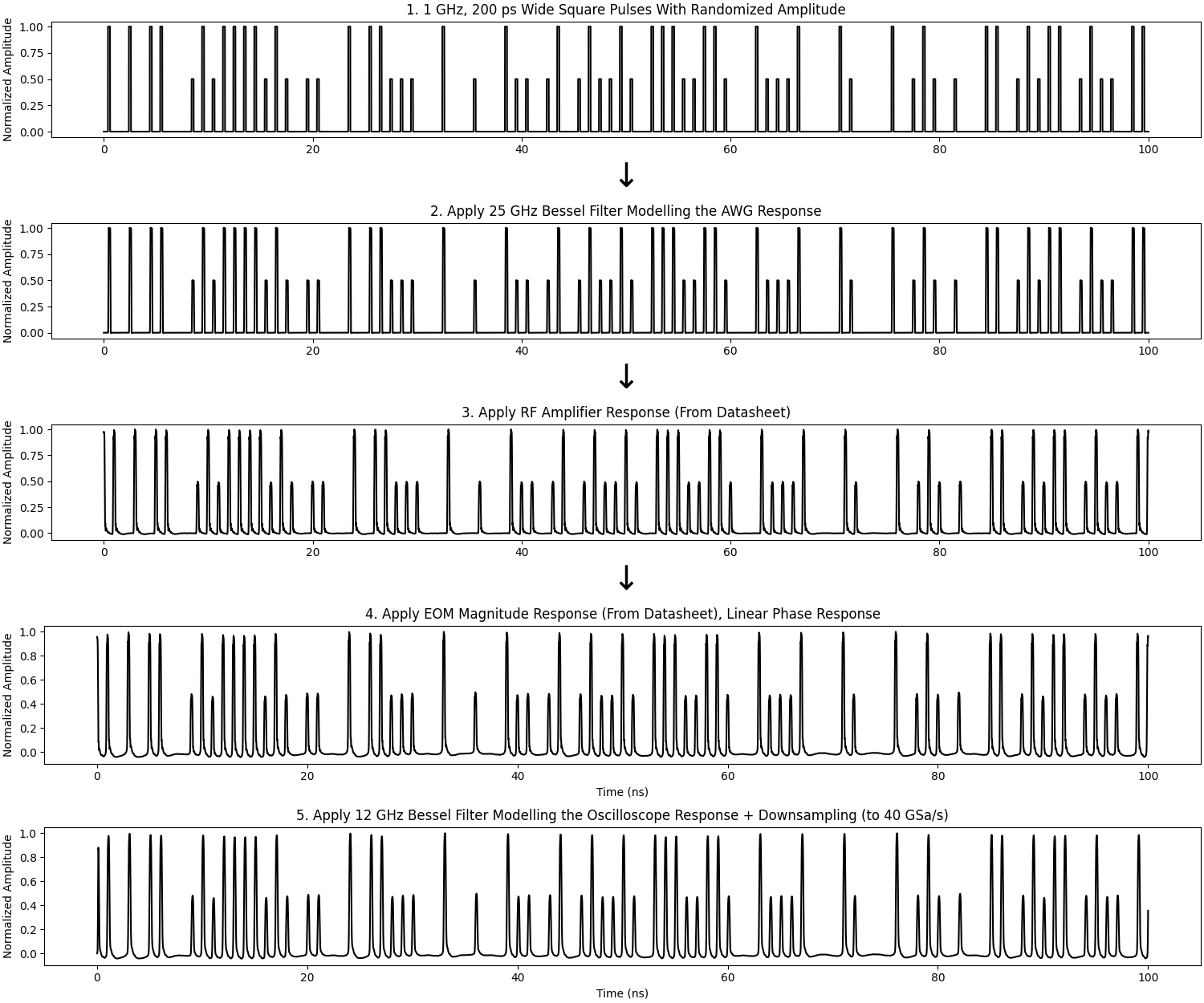}
\caption{Process of simulating actively phase modulated light.}
\label{fig:simulated-pulses}
\end{figure}

To match the experimental characterization, 150 square pulse trains, each with a repetition rate of 1 GHz and a pulse width of 200 ps were generated computationally and fed to the simulator described in Section IIB in the main text (and illustrated in Figure \ref{fig:simulated-pulses}). Each pulse train consisted of 100 pulses with randomized amplitude. These amplitudes correspond to one of the three desired phase changes in a three-state protocol (0,$\pi/2$, and $\pi$). The same procedure was repeated at 500 MHz, 2 GHz, and 3 GHz. For a given repetition rate, the simulated waveforms were normalized using the
maximum amplitude from all the simulated waveforms at that repetition rate. The normalized amplitude corresponds to a phase change (0 corresponding to $0$ and 1 corresponding to $\pi$).

\section{Details Regarding the Experimental Methods for Characterizing Intensity Correlations in the Partially Passive Source}
\label{appendix:intensity_correlations}

The experiment employed the same setup as shown in Figure \ref{fig:characterization-setup}. 100 voltage pulse trains with an amplitude of $V_{\pi}$, each with random multiples of 1 ns between adjacent pulses were generated by the AWG, amplified by the RF amplifier, and sent to the IM. Each pulse train was pseudorandom, consisting of a repeating set of pulses (each set was 100 ns long). For each of the 100 pulse trains, the corresponding intensity modulated pulse train was measured using the photodiode and oscilloscope, after being amplified using the optical amplifier. The entire 100 ns pattern was obtained each time. As the pulse trains consisted of repeating sets, the averaging mode could be used on the oscilloscope to average out the effect of short-term noise. An averaging mode of 1024 was observed to be sufficient for removing short-term noise.

Using the 100 collected optical pulse trains, the goal was to determine how the intensity of a pulse differs depending on the location of the preceding pulse (quantified in ns by $l$). All the occurrences of each case were combined into a mean intensity and standard deviation. The tabulated results can be seen in Table IV in the main text.

\end{document}